# Design of Flexible Meander Line Antenna for Healthcare for Wireless Medical Body Area Networks

Shahid M Ali [A], Cheab Sovuthy [A], Sima Noghanian [B], Qammer H. Abbasi [c], Tatjana Asenova [D], Peter Derleth [D], Alex Casson [E], Tughrul Arslan [F], and Amir Hussain [G]

*Abstract*—a flexible meander line monopole antenna (MMA) is presented in this paper. The antenna can be worn for on-and off-body applications. The overall dimension of the MMA is 37 mm × 50 mm × 2.37 mm³. The MMA was manufactured and measured, and the results matched with simulation results. The MMA design shows a bandwidth of up to 1282.4 (450.5) MHz and provides gains of 3.03 (4.85) dBi in the lower and upper operating bands, respectively, showing omnidirectional radiation patterns in free space. While worn on the chest or arm, bandwidths as high as 688.9 (500.9) MHz and 1261.7 (524.2) MHz, and the gains of 3.80 (4.67) dBi and 3.00 (4.55) dBi were observed. The experimental measurements of the read range confirmed the results of the coverage range of up to 11 meters.

*Clinical Relevance*— Wireless Body Area Network (WBAN) technology allows for continuous monitoring and analysis of patient health data to improve the quality of healthcare services.

## I. INTRODUCTION

Currently, wearable electronic devices have gained popularity among many users for tracking biological data using wireless body area network (WBAN) technology. Their applications might be in healthcare, the military, sports, and electronic gaming (Figure 1) [1]. These applications demand a simple integration of flexible wireless electrical devices into clothing and other wearable devices; therefore, allowing the wearer to use and interact with a variety of devices. The antenna is a key component in a WBAN system. The wearable antenna should fulfil certain requirements, for example, low profile, planar structure, lightweight, flexibility, ease of integration with an electronic device to ensure reliable communication as well as the comfort of the users. A dual-band antenna is preferred for these applications. Various dual-band antennas such as rectangular patch, circular patch, rectangular ring, and coplanar waveguide (CPW)-fed monopole have been investigated for WBAN systems [2]. Some of the proposed directional antennas use electromagnetic bandgap (EBG) structures and artificial magnetic conductor surfaces (AMC) to reduce the specific absorption rate (SAR). In [3], a fractal antenna with meandering slits is proposed. The slits lengthen the path of current and therefore reduce the size. To minimize the size of a microstrip patch, a substrate integrated waveguide (SIW) can be used. Half-mode or quarter-mode SIW are used in dual-band microstrip antennas, reducing the antenna size by 50 % to 75 %, respectively [1]. In [4-5], the antenna footprint is large. Another technique of miniaturization of the antenna is using lumped components, for example, chip inductors or capacitors. In this method, the resonance frequencies can be controlled and pushed to lower frequencies by the inductance or capacitance, without additional physical dimensions.

In this paper, we propose a flexible dual-band MMA with symmetrical inverted slots, and we investigate its operation when it is worn by various subjects in bending and wet situations, as well as in dual bands with different body sizes, where a wearable design is estimated to work for on- and off-body wireless communication. Because of its small size, the results are unaffected by the human body, and when a 10 mm distance is maintained, only minor detuning is found at both bands. This manuscript is organized as follows: following an introduction; however, Section 2 discusses the design steps of the flexible wearable MMA. The results and discussions of the MMA design are provided in Section 3. Finally, Section 4 discusses the conclusion and future direction.

[1]* Research supported by the UK EPSRC COG-MHEAR programmed grant (Grant no. EP/T021063/1)

A. Shahid M Ali and Cheab Sovuthy are with Department of Electrical and Electronic Engineering, Universiti Teknologi, PETRONAS, Malaysia; shahid_17006402@utp.edu.my; sovuthy.cheab@utp.edu.my

B. Sima Noghanian is with CommScope; sima_noghanian@ieee.org

C. Qammer H. Abbasi is with the James Watt School of Engineering, University of Glasgow, UK; Qammer.Abbasi@glasgow.ac.uk

D. Tatjana Asenova and Peter Derleth are with Sonova AG, Switzerland; tatjana.asenov@sonova.com , peter.derleth@sonova.com

E. Alex Casson is with the University of Manchester; UK, alex.casson@manchester.ac.uk

F. Tughrul Arslan is with the University of Edinburgh, UK; T.Arslan@ed.ac.uk

G. Amir Hussain is with the School of Computing, Edinburgh Napier University, UK; A.Hussain@napier.ac.uk

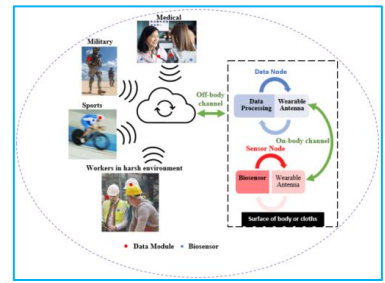

Figure 1. Healthcare design in WBAN system.

## II. DESIGN TOPOLOGY

Figure 2 illustrates the antenna structure (MMA). All simulations in this study were done by CST Microwave Studio (CST MWS). The MMA is composed of a planar structure with symmetrical inverted slots and a truncated shape ground fabricated of ShieldIt™ and copper foil, thicknesses (t in mm) of 0.17 and 0.035, respectively, on a textile felt. The dielectric constant (ε$_r$) of this felt is 1.3, its loss tangent (tan δ) is 0.044, and thickness (t in mm) is 1.5. Under the ground plane, a denim textile with ε$_r$ = 1.43, tan δ of 0.02, and a thickness (t in mm) of 0.7 were used to produce the isolation between the MMA and the body skin. Therefore, the MMA is excited through a microstrip line, which is added to a stripline ($L_{mon}$, in mm) of 40.20. Therefore, adding slots provide a loading effect, and as a result, the MMA quality factor (Q) and size are reduced. The equations described in [6] were used to calculate the MMA dimensions. The dimensions of the final design of MMA are provided in Table 1.

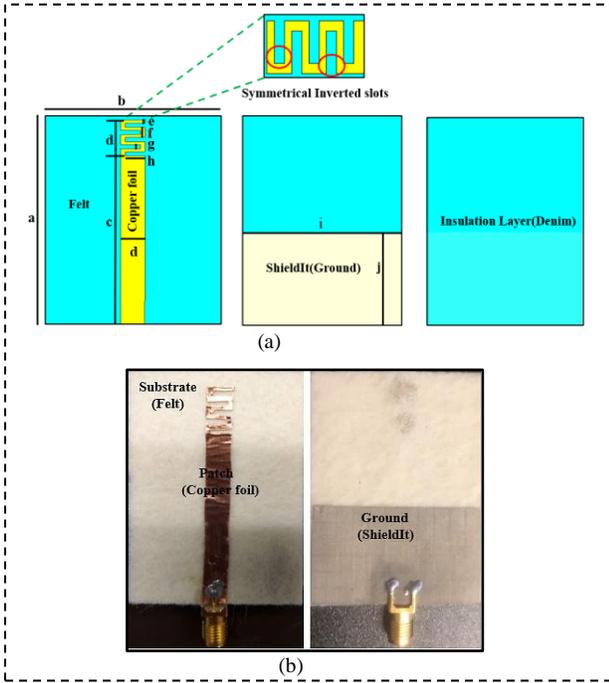

(a)

(b)

Figure 2. (a) MMA different views and (b) the fabricated design.

TABLE 1. The dimensions of the MMA design.

| Symbols | Dimensions (mm) |
|---|---|
| a | 50.00 |
| b | 37.20 |
| c | 40.20 |
| d | 8.49 |
| e | 0.73 |
| f | 2.38 |
| g | 1.00 |
| h | 14.13 |
| i | 37.20 |
| j | 22.00 |

III. RESULTS AND DISCUSSION

To find the optimum design parameters, a study was conducted. All the simulations and the measurements were done within the frequency range of 1 – 7 GHz.

*A. Simulation Results*

In the simulation, CST Microwave Studio® (CST MWS) was used for evaluating the proposed MMA performance as well as performing a parametric tolerance analysis.

*A. On-Body Communication System*

To evaluate the MMA, it is vital to investigate its characteristics while positioned on the various parts of the human body, for example, the chest and arm. The 3D body model, along with a volume of 200 × 200 × 50 mm³ and various thicknesses (in mm) such as skin (4), fat (8), and muscle (40), was utilized. Further, the muscle was assumed to have ε$_r$ of 52.7 and 48.2, as well as a conductivity (σ) of 1.95 and 6 S/m, in the higher and lower operating bands, respectively.

Similarly, a cylindrical arm (in mm) with a diameter of 50 and a length of 150 was selected using the method described in [7-8]. The air gap (in mm) between the MMA and the body was assumed to be 3, 5, or 10. This air-gap height was chosen to illustrate the use of similar thicknesses (t) of numerous textiles, which are similar to an air-gap because the relative permittivity of textile is close to that of air. The results showed that the MMA was almost decoupled from the body at 10 mm, and this gap was assumed for the rest of the study. The simulated S$_{11}$ at various air-gaps is shown in Figure 3.

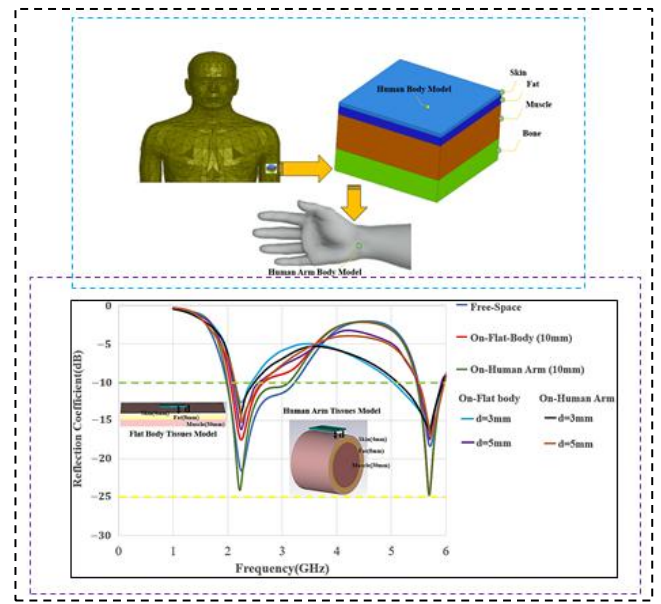

Figure 3. S$_{11}$ in various cases at several positions from the body.

*B. Measurement Results*

The S-parameters were measured with a VNA (PNA-X N5242A), and calibrated by the calibration module (NA4691B), which are discussed in [9].

*1. Reflection Coefficient and Bandwidth*

The S$_{11}$ was measured in three positions that are shown in Figure 4: in free space, on the chest, and on the arm. S$_{11}$ was less than -10 dB in both bands. When tested on different weights (65kg and 110kg), therefore, the MMA performance on the body was found to be comparable to that in the free space. The performance of the MMA in bending situations was then examined. The MMA was tested on the polystyrene cylinders with diameters (D in mm) of 100, 80, 70, 60, and 50 with a ε$_r$ of ~1. In these cases, there was a minor change in the resonance frequency, as shown in Figure 5. The results show the MMA functions with minimum frequency detuning within the operating band.

Finally, we tested the condition where the MMA was dampened by soaking the felt textile in water until it was saturated with water (Figure 6). The damping had a significant impact on $S_{11}$. When the MMA was tested after an hour. The evaporation of water from the substrate caused some changes, but the values were not back to the normal condition. The $S_{11}$ values after 2 hours were similar to those after one hour. When the MMA was completely dried, at a room temperature of 26 °C and 30 % air humidity, $S_{11}$ values were nearly back to those obtained before dumping the antenna.

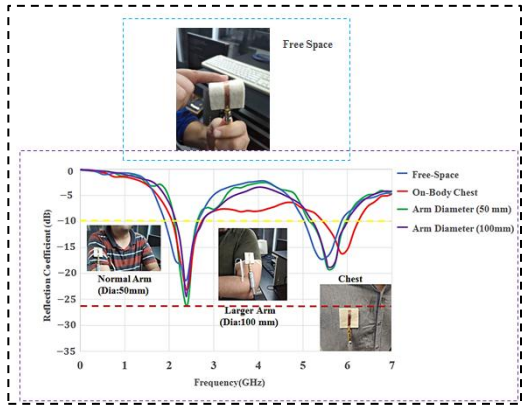

Figure 4. $S_{11}$ tested in various situations such as (chest and arm: 50, 100 (diameter in mm).

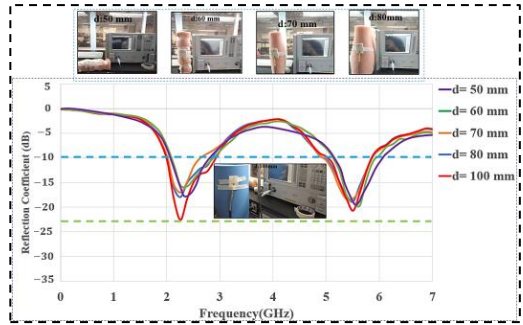

Figure 5. $S_{11}$ on different cylindrical diameters.

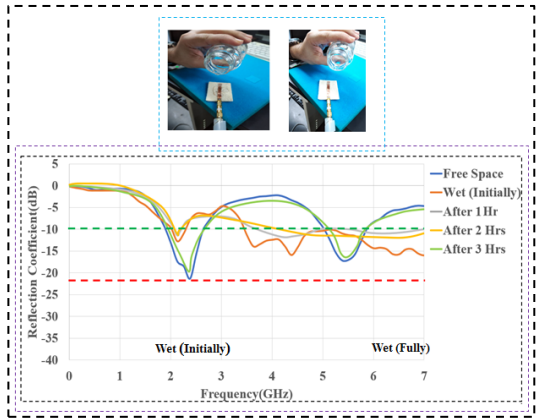

Figure 6. $S_{11}$ in wet conditions (free space, different wet conditions).

## A. 3D Radiation Patterns

The far-fields were tested inside an anechoic chamber. The MMA design was mounted on a stand that served as a receiving antenna, while a horn antenna was a transmitting antenna and rotating in elevation and azimuth planes. The VNA was connected to the MMA to measure the far-field on the body models. The MMA was tested on chest and arm models. The values of their parameters are similar to those mentioned in the simulation for both the lower and upper bands, respectively [9]. As shown in Figure 7, omnidirectional radiation patterns were obtained in both bands, which show the MMA is a suitable candidate for WBAN systems.

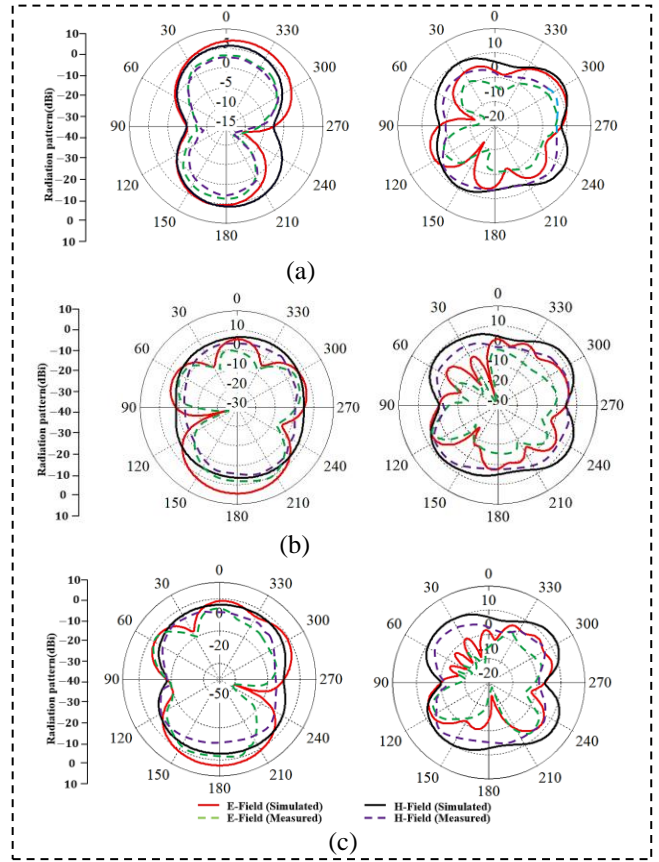

Figure 7. Measured far-field patterns (lower band, left, and upper band, right), (a) free-space at 2.2/5.6 GHz), (b) chest at 2.3/5.7 GHz), (c) arm at 2.2/5.7 GHz.

## 3. On-body Path-loss Calculation

The path-loss ($S_{21}$) measurement ensures a sufficient link budget for on-body communications. We calculated $S_{21}$ for numerous situations of free-space in an open environment (e.g. parking lot), as demonstrated in Figure 8. Both Tx and Rx MMAs were mounted on tripods with a suitable height of one meter and vertically aligned. The distance (r in m) was set at 2, 5, 8, and 11.

Similarly, for on-body communication, $S_{21}$ values were measured. As indicated in Figure 9, the MMAs were put on the chest and the arm. The results show that MMAs provide similar $S_{21}$ values in an open space as compared to the lossy body in both bands. The average path-loss exponent was found to be $n = 3$, and it matches the urban areas using equation (1).

$$PL_{dB}(r) = PL(r_0) + 10n\log\left(\frac{r}{r_0}\right), \quad (1)$$

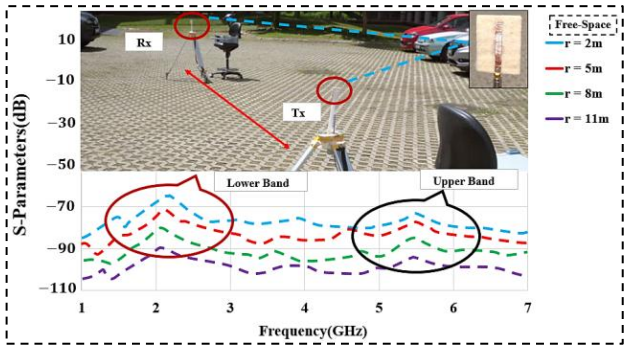

Figure 8. $S_{21}$ in free space at various distances between the Tx/Rx MMAs (r: 2 to 11 m).

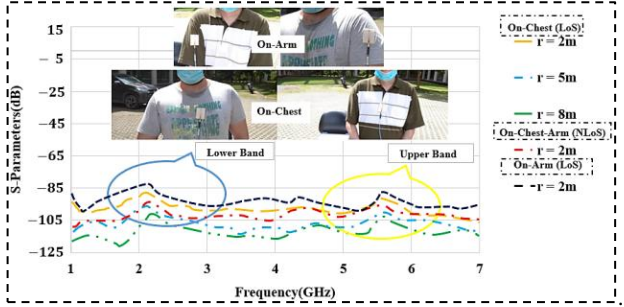

Figure 9. $S_{21}$ on the body locations in LOS and NLoS between Tx/Rx MMAs (r: 2 to 8 m).

*3.3. Specific Absorption Rate (SAR)*

SAR plays an important role in ensuring safety [9-11]. It depicts how tissues absorb radiations, and as a result, it increases the body temperature. For 1g-tissue (Federal Communication Commission (FCC) U.S. standard, average SAR must be less than 1.6 W/kg, or it must be less than 2 W/kg if using 10-g average is used according to ICNIRP, European standard. Figure 10 shows the simulated SAR maps for the arm and chest under the premise of the MMA with a maximum of 20 dBm input power.

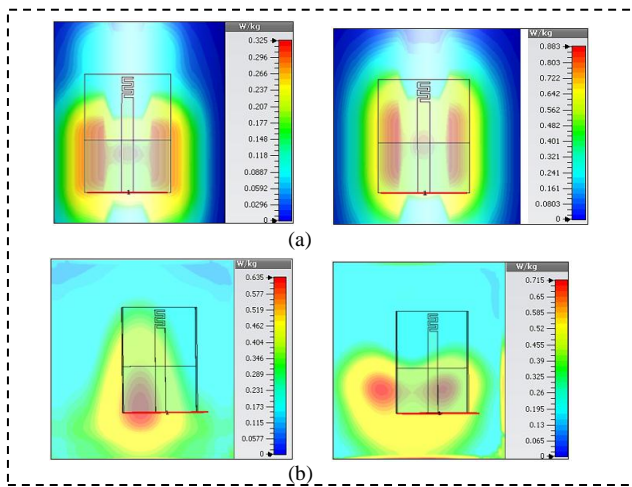

Figure 10. SAR (10-g): (a) arm (at 2.2 (left) and 5.7 GHz (right)), and (b) on the chest (2.3 (left) and 5.7 GHz (right)).

## IV. CONCLUSION

In this work, a flexible MMA with a wide bandwidth of 2.2–3.0 GHz and 5.6–6.0 GHz was designed and manufactured. The MMA performance was investigated in numerous situations, including in free space and when placed on the body (chest and arm), with various body sizes, bending, and damp situations. The MMA was manufactured, and the measured results matched those from the simulation. In free space as well as on the body, the MMA provided a wide BW, omnidirectional radiation patterns, and suitable gains. Therefore, the MMA platform has the potential to be used in next-generation multi-modal hearing-aid designs [12].

## REFERENCES

[1] S. M. Ali et al, "Recent advances of wearable antennas in materials, fabrication methods, designs, and their applications: State-of-the-art," *Micromachines*, vol. 11, no. 10, 2020.

[2] T. T. Le and T. Y. Yun, "Miniaturization of a dual-band wearable antenna for WBAN applications," *IEEE Antennas Wirel. Propag. Lett.*, vol. 19, no. 8, pp. 1452–1456, 2020.

[3] A. Arif et al, "A compact, low-profile fractal antenna for wearable on-body WBAN applications," *IEEE Antennas Wirel. Propag. Lett.*, vol. 18, no. 5, pp. 981–985, 2019.

[4] S. Agneessens and H. Rogier, "Compact half diamond dual-band textile HMSIW on-body antenna," *IEEE Trans. Antennas Propag.*, vol. 62, no. 5, pp. 2374–2381, 2014.

[5] X. Q. Zhu et al, "A compact dual-band antenna for wireless body area network applications," *IEEE Antennas Wirel. Propag. Lett.*, vol. 15, 98–101, 2016

[6] C. C. Hsu and H. H. Song, "Design, Fabrication, and characterization of a dual-band electrically small meander-line monopole antenna for wireless communications," *Int. J. Electromagn. Appl.*, vol. 3, no. 2, pp. 27–34, 2013.

[7] S. Yan and G. Vandenbosch, "Design of wideband button antenna based on characteristic mode theory," *IEEE Trans. Biomed. Circuits Syst.*, vol. 12, no. 6, pp. 1383–1391, 2018.

[8] "Body tissue dielectric parameters, radiofrequency safety," 2015.

[9] S. M. Ali *et al.*, "Design and evaluation of a flexible dual-band meander line monopole antenna for on- and off-body healthcare applications," *Micromachines*, vol. 12, no. 5, 2021.

[10] P. J. Soh, et al, "Specific Absorption Rate (SAR) evaluation of textile antennas," *IEEE Antennas Propag. Mag.*, vol. 57, no. 2, pp. 229–240, 2015.

[11] W. Wang, et al, "A low-profile dual-band omnidirectional Alford antenna for wearable WBAN applications," *Microw. Opt. Technol. Lett.*, vol. 62, no. 5, pp. 2040–2046, 2020.

[12] A. Adeel, J. Ahmad, H. Larijani and A. Hussain. "A Novel Real-Time, Lightweight Chaotic-Encryption Scheme for Next-Generation Audio-Visual Hearing Aids". *Cogn Comput,* 12, 589–601 (2020).